# Cavity ring-down spectroscopy of $CO_2$ near $\lambda$ = 2.06 μm: Accurate transition intensities for the Orbiting Carbon Observatory-2 (OCO-2) "strong band"

Hélène Fleurbaey, Hongming Yi[1], Erin M. Adkins, Adam J. Fleisher*, Joseph T. Hodges

*Material Measurement Laboratory, National Institute of Standards and Technology, 100 Bureau Drive, Gaithersburg, MD 20899, USA*

Dated: 5/19/2020

[1] Present address: *The Department of Civil and Environmental Engineering, Princeton University, E208 Engineering Quadrangle, Princeton, NJ 08544, USA*

*Corresponding author: adam.fleisher@nist.gov

## Abstract

The $\lambda = 2.06$ μm absorption band of $CO_2$ is widely used for the remote sensing of atmospheric carbon dioxide, making it relevant to many important top-down measurements of carbon flux. The forward models used in the retrieval algorithms employed in these measurements require increasingly accurate line intensity and line shape data from which absorption cross-sections can be computed. To overcome accuracy limitations of existing line lists, we used frequency-stabilized cavity ring-down spectroscopy to measure 39 transitions in the $^{12}C^{16}O_2$ absorption band. The line intensities were measured with an estimated relative combined standard uncertainty of $u_r = 0.08$ %. We predicted the *J*-dependence of the measured intensities using two theoretical models: a one-dimensional spectroscopic model with Herman-Wallis rotation-vibration corrections, and a line-by-line *ab initio* dipole moment surface model [Zak et al. JQSRT 2016;177:31-42]. For the second approach, we fit only a single factor to rescale the theoretical integrated band intensity to be consistent with the measured intensities. We find that the latter approach yields an equally adequate representation of the fitted *J*-dependent intensity data and provides the most physically general representation of the results. Our recommended value for the integrated band intensity equal to $7.183 \times 10^{-21}$ cm molecule$^{-1}$ ± $6 \times 10^{-24}$ cm molecule$^{-1}$ is based on the rescaled *ab initio* model and corresponds to a fitted scale factor of $1.0069 \pm 0.0002$. Comparisons of literature intensity values to our results reveal systematic deviations ranging from −1.16 % to +0.33 %.

# 1. Introduction

Increasing carbon dioxide ($CO_2$) concentrations caused by anthropogenic emissions [1] are a primary contributor to the estimated global warming of 1 °C above pre-industrial levels [2]. To understand the consequences of this trend, climate models that rely on accurate knowledge of the global carbon cycle are required. Therefore, adequate constraints on the spatial and temporal variability of $CO_2$ fluxes are necessary.

Satellite measurements can provide the information needed to constrain $CO_2$ sources and sinks on time and spatial scales relevant to climate modeling with validation provided by ground-based carbon-monitoring networks, such as TCCON. The GOSAT, GOSAT-2 (JAXA), OCO-2 and OCO-3 (NASA), TanSat (CAS), and the upcoming MicroCarb (CNES) missions all use satellite-based spectrometers to analyze reflected sunlight within the $CO_2$ bands at 1.61 μm and 2.06 μm. These instruments report the atmospheric-column-integrated $CO_2$ mole fraction ($XCO_2$) and rely on first-principles, line-by-line radiative transfer models to account for the dependence of the absorption cross-section on wavelength, pressure, temperature, and air-composition. Absorption cross-sections are frequently calculated offline and stored in look-up tables (i.e. ABSCO [3]) to increase the speed of retrieval algorithms.

Importantly, the carbon dioxide absorption cross-sections are scaled by the intensities of the lines comprising these bands. Because spectroscopy is only part of the retrieval error budget, line intensities need to be known to a higher accuracy and precision than the $XCO_2$ relative standard uncertainty target, which is 0.3 % for the OCO-2 mission [4]. Only recently have line intensities of such accuracy been reported for $CO_2$ [5], and the demand for improved accuracy continues to generally motivate recent advances in related cavity-enhanced methods (e.g., [6, 7]). Beyond single lines, it is highly desirable to minimize uncertainty in the intensities of many lines to establish reliable spectroscopic constraints and band-to-band consistency when retrieving $CO_2$ concentration from observed spectra [8]. Because systematic errors in spectroscopic parameters can bias regional $XCO_2$ retrievals and flux estimates, reference-quality measurements of $CO_2$ spectroscopic parameters are necessary to achieve these targets.

ABSCO v5.0 OCO-2 retrieval algorithms use line lists provided by Devi et al. [9] and Benner et al. [10] for the 1.61 μm and 2.06 μm bands, respectively. Oyafuso et al. [3] evaluated ABSCO v5.0 single-band $XCO_2$ retrievals against TCCON retrievals, which have an estimated relative uncertainty of 0.2 % [11]. Both bands showed a clear discrepancy with the TCCON $XCO_2$ retrievals. A joint theoretical and experimental investigation [12] into the 1.61 μm band line intensities reported agreement between the *ab initio* calculations and cavity ring-down spectroscopy measurements within 0.3 % and an offset between the experimental measurements and the Devi work of 1.4 %. As noted by Oyafuso et al. [3], scaling the ABSCO v5.0 absorption coefficients by this factor would bring the 1.6 μm band $XCO_2$ estimates into much better agreement with the TCCON-reported values that were calibrated against *in situ* profile measurements. At the time of the Oyafuso article publication, Benner et al. [10] was the only source of reference intensities for the 2.06 μm band.

In this paper, we present new measurements of transition intensities in the $20013 \leftarrow 00001$ rotation-vibration band of $^{12}C^{16}O_2$, realized using frequency-stabilized cavity ring-down spectroscopy (FS-CRDS). These measurements are compared to *ab initio* calculations [13] and serve as reference intensity measurements for the 2.06 μm band. Because of its higher intensity relative to the 1.61 μm band, the 2.06 μm band studied here is colloquially referred to as the "strong band". Figure 1 shows the calculated absorbance of this strong band and the lines included in this study. The spectra were recorded in two separate sessions with slightly different experimental configurations, yielding two data sets A and B, recorded respectively in November-December 2018 and May 2019. Data set A was acquired with a suite of narrowband distributed feedback (DFB) lasers and included only the 24 lines labeled in red in Fig. 1. Data set B was acquired with a broadly tunable external cavity diode laser (ECDL), enabling measurement of all 39 labeled lines. Both sets were used for the final analysis.

The remainder of the paper is organized as follows. The next section gives an overview of the experimental setup and the data recording procedure. Data analysis is then described, and methods for calculating the integrated band intensity from the line-by-line analysis are discussed. Specifically, we compared two models for line-by-line intensities within the $20013 \leftarrow 00001$ vibrational band: a standard spectroscopic model and an *ab initio* quantum chemistry model. Finally, the results are discussed in comparison to the literature values, including available line lists, databases, and prior experiments.

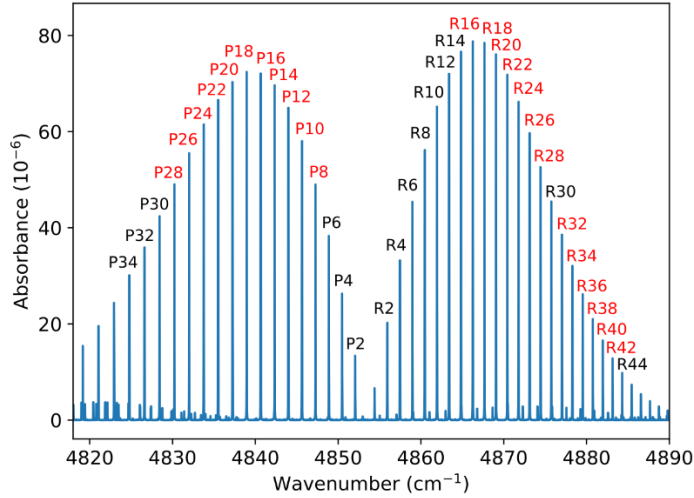

**Fig. 1.** Calculated absorbance spectrum of $CO_2$ near $\lambda = 2.06$ μm (pressure $p = 6.7$ kPa, pathlength $L = 75$ cm, $CO_2$ mole fraction $\chi_{CO_2} = 50$ μmol/mol, line parameters from HITRAN2016 [18]). The $^{12}C^{16}O_2$ transitions labeled in red were recorded in both data sets, whereas those labeled in black were only recorded in data set B (2019).

## 2. Experimental setup

Previous implementations of the cavity ring-down spectrometer used herein, operating at $\lambda \approx 2$ μm, were described in Refs. [14, 15]. The setup comprises a high-finesse length-stabilized optical cavity, a gas flow system, a probe laser, and detection apparatus, as well as a computer with a digitizer and analysis software. The spectrometer leverages the FARS (frequency-agile, rapid scanning) [16] method to scan the probe laser frequency in the vicinity of a molecular line. Figure 2 presents a simplified diagram of the experimental setup.

The optical cavity was created by two high-reflectivity, multi-wavelength-coated mirrors of power reflectivity $R = 0.99994$ at $\lambda = 2$ μm and $R \approx 0.95$ at $\lambda = 633$ nm separated by invar rods resulting in a mirror separation of nominally 75 cm. Because the mirror coatings were designed to have maximum reflectivity at a wavelength of $\lambda = 2$ μm ($\tilde{\nu} = 5000$ cm$^{-1}$), the base losses varied over the studied wavenumber region ($\tilde{\nu} = 4824$ to $4885$ cm$^{-1}$) from $2 \times 10^{-4}$ to $6 \times 10^{-4}$ (decay time constant in baseline from 13 to 4 μs). One of the mirrors was mounted to a piezoelectric transducer to actively stabilize the cavity length by maintaining resonance with a reference HeNe laser with long-term frequency stability of 1 MHz [17]. The free spectral range ($\nu_{\text{FSR}}$) of this cavity was measured and reported in Ref. [15] to be 200.07 MHz ± 0.03 MHz.

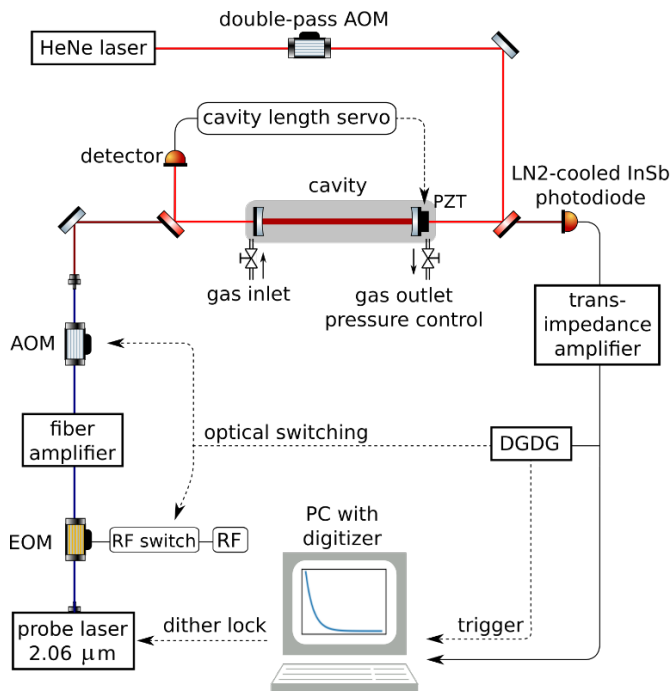

**Fig. 2.** Simplified diagram of the FS-CRDS experimental setup (optical isolators and mode-matching optics are not shown). Acronyms: AOM, acousto-optic modulator. EOM, electro-optic modulator. DGDG, digital gate-and-delay generator. LN2, liquid nitrogen. PC, personal computer. PZT, piezoelectric actuator. RF, radiofrequency.

The gas sample was a NIST-certified gravimetric mixture [18] of petrochemically derived $CO_2$ in $N_2$ with $CO_2$ mole fraction $\chi_{CO_2}$ = 49.826 µmol/mol ± 0.019 µmol/mol. The isotopic composition of $CO_2$ within the certified sample was measured via Fourier transform spectroscopy (FTS) relative to a secondary reference $CO_2$ sample with known value assignments for $\delta^{13}C_{\text{VPDB}}$ and $\delta^{18}O_{\text{VPDB}}$, where VPDB indicates the $\delta$ values are on the Vienna Pee Dee Belemnite scale. Through FTS and the secondary reference sample, the certified sample was assigned $\delta^{13}C_{\text{VPDB}} \approx$ −40 ‰. To compute the relative abundance of the $^{12}C^{16}O_2$ (626) isotopologue, we assumed $\delta^{18}O_{\text{VPDB}} \approx$ −24 ‰ which is typical of petroleum-derived carbon dioxide. Assuming a stochastic distribution of C and O isotopes yields the relative abundance $\chi_{626}$ = 0.98464 ± 0.00007 so that the intensities we report here were normalized to the HITRAN relative abundance value of $\chi_{626,\text{HT}}$ = 0.984204 [19] by multiplying by the factor $\chi_{626,\text{HT}}/\chi_{626}$ = 0.99956 ± 0.00007.

The sample was flowed through the optical cavity after passing through a needle valve, and the intra-cavity pressure was stabilized using a voltage-controlled valve placed between the cavity gas outlet and a vacuum pump. The chosen volumetric flow rates (20 std. cm$^3$ min$^{-1}$ to 50 std. cm$^3$ min$^{-1}$) were high enough to largely eliminate the effects confounding interactions of $CO_2$ with metal surfaces in the ring-down cavity (estimated gas residence time of 60 s), and a pressure stability of 10 Pa (about 0.1 Torr) at each setpoint was achieved over a time interval of approximately 8 h. The sample pressure was accurately measured in two steps. First, pressure was measured at a single point in the flow system (located immediately upstream of the optical cavity) by a resonant silicon manometer with a full-scale range of 130 kPa calibrated against a secondary NIST pressure standard. Second, to measure the flow-induced pressure drop between the sample gas upstream of the manometer and that within the optical measurement volume, a differential pressure measurement was performed between two points in the flow which were symmetrically located about the geometric center of the optical cavity. Analyzed in combination, the absolute and differential pressure measurements resulted in a relative combined type-B uncertainty associated with pressure of 0.02 %. The temperature was measured by a platinum resistance thermometer in good thermal contact with the outside of the metallic sample cell, and temperature stability was enhanced by surrounding the sample cell with a passive thermally insulated enclosure. Again, the temperature probe was calibrated using a secondary NIST standard.

Different probe lasers were used for the two recording sessions (data sets A and B). For data set A recorded in 2018, four different DFB laser diodes were used to measure the transitions labeled in red in Fig. 1. Before the second recording session, the DFBs were replaced by an ECDL with a tuning range from 1975 nm to 2073 nm (data set B). The rest of the optical system remained unchanged.

The fibered output of the probe laser passed through an optical isolator and was coupled into an electro-optic modulator (EOM) which generated sidebands used for FARS scanning. The frequency-modulated beam was then amplified by an optical fiber amplifier with variable gain. The amplified continuous-wave laser subsequently passed through an acousto-optic modulator (AOM) and several Faraday isolators (total optical isolation of ≥ 60 dB) and continued in free space through mode-matching optics before being injected into the optical cavity.

The probe transmission was collected by a liquid-nitrogen-cooled InSb photodetector followed by a transimpedance amplifier ($10^5$ V/A). When the transmission reached a user-defined threshold, a digital gate-and-delay generator (DGDG) triggered the decay acquisition by shuttering the incident light, acting on two optical switches simultaneously: the AOM was switched off and the EOM sideband generation was also switched off by a rapid microwave switch. The resulting decay signals were recorded by a digitizer with a resolution of 16 bits and electronic bandwidth of 125 MHz. Digitizer nonideality was independently characterized for each recording session using the procedure described in Ref. [5].

From the optical power at the photodetector at trigger threshold, and assuming a mirror power loss coefficient from absorption and scattering of $3\times10^{-5}$, we estimate the intracavity intensity to be <0.4 W cm$^{-2}$ with a beam waist radius $\omega_0$ = 0.56 mm — an intensity which is several orders of magnitude less than the saturation intensity [20] predicted for the strongest $CO_2$ line studied here. Therefore, saturation effects at sample pressures >6 kPa were negligible.

Spectra were acquired for each individual line, with several replicates recorded sequentially (5 to 15 replicates for data set A, 3 for data set B) in the following manner. First, the probe laser was tuned to the center of the line and its frequency was finely adjusted to bring one of the EOM sidebands onto resonance with a TEM$_{00}$ mode of the ring-down cavity. At that point, the laser frequency was dither-locked to the cavity with feedback to the laser current. Then the EOM sideband frequency was stepped in increments of $\nu_\text{FSR}$ from −6 GHz to +6 GHz (where a negative frequency implies that the negative sideband was in resonance with the cavity), resulting in a 60-point scan covering a 12 GHz span. At each frequency point, nominally 100 (set A) or 300 (set B) decay events were averaged. A summary of experimental conditions for data sets A and B is shown in Table 1.

We note that only one value of the pressure and temperature was recorded for each individual FARS scan (90 s scan time). However, the temperature appeared to be rising continuously throughout the day (~8 h) corresponding to a change between 1 mK and 5 mK during any given scan. On the time scale of a scan, the pressure variation was below the digital resolution of the gauge (~1.3 Pa, or 0.01 Torr).

Table 1

Summary of experimental conditions for the two data sets comprising transition intensities ($S$) for the 20013 ← 00001 $^{12}C^{16}O_2$ band near $\lambda = 2.06$ μm.

| Description | Data set A 2018 | Data set B 2019 |
|---|---|---|
| Dates recorded | 14 Nov. to 4 Dec. | 6 May to 14 May |
| Number of $^{12}C^{16}O_2$ transitions | 24 | 39 |
| Probe laser(s)[a] | DFB lasers (quantity 4) | ECDL |
| Probe laser linewidth[b] | <2 MHz | 200 kHz |
| Digitizer nonideality[c] | $\tau_{corr} = b_0 + b_1\tau$ $b_0 = 0.01078$ $b_1 = 1.00175$ | $\tau_{corr} = b_0 + b_1\tau + b_2\tau^2$ $b_0 = 0.009456$ $b_1 = 1.002414$ $b_2 = 6.558 \times 10^{-5}$ |
| Number of decays averaged | 100 | 300 |
| FS-CRDS precision ($\sigma_\tau/\tau$) | 0.1 % to 0.5 % | 0.04 % at 4885 cm$^{-1}$ to 0.12 % at 4825 cm$^{-1}$ |
| DGDG trigger threshold | Variable, 0.2 V to 1.8 V | 2 V |
| Number of replicates per pressure | 5 to 15 | 3 |
| Flow rate[d] | 20 std. cm$^3$ min$^{-1}$ | 50 std. cm$^3$ min$^{-1}$ |
| Pressure-gradient correction | None | Two-point measurement −0.17 % at 6.7 kPa −0.02 % at 20 kPa |
| Observed temperature range | 25 °C to 25.6 °C | 25.4 °C to 26 °C |
| Baseline etalon | None fitted | 2.82 GHz period |
| Line profile[e] | SDNGP with fixed $a_w = 0.08$[f] | SDNGP with constrained $a_w$ |
| Line profile bias correction[g] | −0.049 % | −0.036 % |

[a] DFB, distributed feedback. ECDL, external cavity diode laser.
[b] Manufacturer's specification. Integration time: DFB, none provided; ECDL, 50 ms.
[c] Digitizer settings: Vertical range = 10 V; Impedance = 50 Ohm; Decay fit window = variable, from 1.5 μs to 10τ.
[d] Readout from an uncalibrated flow gage.
[e] SDNGP, speed-dependent Nelkin-Ghatak profile.
[f] $a_w = \Gamma_2/\Gamma_0$, where $\Gamma_2$ and $\Gamma_0$ are the speed-dependent and velocity-averaged collisional relaxation rates [22-25].
[g] Bias correction relative to the chosen reference profile, the Hartmann-Tran profile with $\eta \leq 0.5$.

## 3. Data analysis

The total fractional optical power loss per unit length in the optical cavity is given by the relation

$$L_{tot} = \frac{1}{c\tau} = L_0 + \alpha, \tag{1}$$

where $c$ is the speed of light, $\tau$ is the decay time constant, $L_0$ is the baseline loss divided by the cavity length (mirrors, parasitic etalons, etc.) and $\alpha$ is the absorption coefficient of the medium inside the cavity.

Measurements of digitizer nonidealities mentioned in the previous section resulted in a systematic correction of the recorded decay time constants by one of the following two relations: $\tau = b_0 + b_1\tau_a$ (data set A) or $\tau = b_0 + b_1\tau_a + b_2\tau_a^2$ (data set B), where $\tau_a$ is the raw "apparent" time constant and the coefficients $\{b_0, b_1\}$ and $\{b_0, b_1, b_2\}$ were determined through the traceable procedure described in Ref. [5]. For data set A, digitizer nonidealities were measured using an arbitrary waveform generator (AWG) as a transfer between the digitizer under test and a metrology-grade reference digitizer, whereas for data set B, coefficients were determined directly from the unamplified AWG output which we confirmed to be consistent with the reference board. The two measurements of nonideality performed nearly 6 months apart agreed to within 0.05 %.

In both data sets, lines were recorded for 5 pressure values between 6.7 kPa and 20 kPa [(50, 75, 100, 125, 150) Torr equal to (6.7, 10, 13.3, 16.7, 20) kPa]. To account for changes in the effective cavity length with pressure, the frequency spacing between spectral data points was calculated from the empty-cavity $\nu_{FSR}$ and the known dependence of the refractive index on gas density [21].

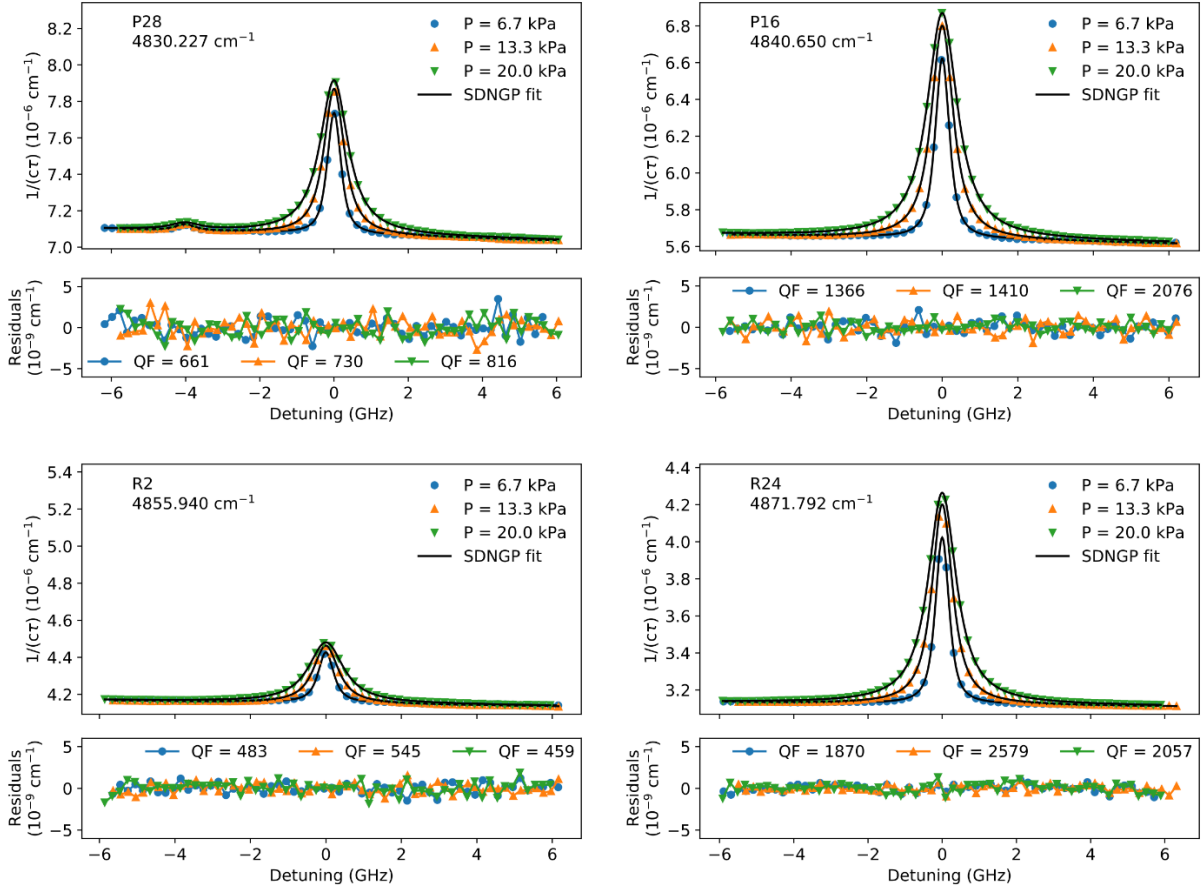

**Fig. 3.** Sample spectra of the lines P28, P16, R2 and R24, recorded at pressures of 6.7 kPa, 13.3 kPa and 20.0 kPa in data set B. In this example, each spectrum was fitted with the speed-dependent Nelkin-Ghatak profile (SDNGP) and the resulting residuals are shown below each spectrum along with the corresponding quality factor (QF). The baseline noise amplitude scaled with the base losses and was larger in the P-branch. An etalon of period 2.82 GHz was included in the fits, and the fitted etalon amplitude also scaled with the base losses. Note that the relative vertical scale is the same for the four represented transitions.

Data analysis for each of the two data sets was performed independently. A representative analysis procedure is described here. In the first step, the spectra within each respective data set were fitted individually with a chosen line profile (Fig. 3). For a recent review of relevant line profiles and their parameters, see Ref. [22]. For data set A, the chosen line profile was a speed-dependent Nelkin-Ghatak profile (SDNGP) [23] with a fixed value for the dimensionless speed-dependent broadening parameter of $a_w \equiv \Gamma_2/\Gamma_0 = 0.08$ (where $\Gamma_2$ and $\Gamma_0$ are the speed-dependent relaxation rate and the velocity-averaged Lorentzian halfwidth, respectively) [24, 25]. For data set

B, four different line profiles were used: Galatry (GP) [26], Nelkin-Ghatak (NGP), speed-dependent Voigt (SDVP), and SDNGP. For this last profile, the dimensionless speed-dependent parameter $a_w$ was fixed to a value calculated for each line as

$$a_w = (1-n)\frac{2}{3}\frac{m_p/m_a}{1+m_p/m_a}, \qquad (2)$$

where $n$ is the temperature exponent of the collisional broadening coefficient, and $m_p/m_a$ is the perturber-to-absorber mass ratio [25]; this value of $a_w$ varied between 0.057 and 0.08 for the different lines. In all cases, floated parameters included line position ($\nu_0 + \Delta$, where $\Delta$ is the pressure shift), integrated spectral area ($A(T)$), collisional broadening halfwidth ($\Gamma_0$), Dicke narrowing frequency ($\nu_{VC}$), and the slope and intercept of a linear background. The Doppler halfwidth ($\Gamma_D$) was fixed to the theoretical value based on the measured temperature, absorber mass and transition frequency.

In practice, to the extent that the chosen line profile differs from the actual line shape, the retrieved area of a transition depends on the line profile used to fit the spectral data. Indeed, systematic differences were observed in data set B between the areas retrieved using different profiles; areas obtained with NGP were typically about 0.1 % to 0.2 % lower than those obtained from SDVP, with the two other profiles falling in between those two extreme cases. Because no systematic structure was observed in the residuals, it was hard to justify which profile best represented the experimental results. To be relatively consistent with the analysis of data set A, we chose for data set B to report the line areas fitted by the most sophisticated profile: the SDNGP with constrained $a_w$.

We evaluated bias in the choice of the line profile by fitting to a series of simulated reference profiles. Here, the reference profile is the Hartmann-Tran profile (HTP) [22] with values for the parameter describing partial correlation between velocity and rotational state changes due to collisions, $\eta$, within the range of 0 to 0.5. The upper-bound of $\eta = 0.5$ was determined empirically by establishing the value of $\eta$ at which the quality factor (QF) of the fit dropped below the experimental signal-to-noise ratio. Fits to the simulated reference profiles using appropriate spectroscopic parameters for the $CO_2$ band studied here resulted in estimates of the systematic bias for both data sets: data set A bias correction, −0.049 %; data set B bias correction, −0.036 %, relative to the chosen HTP reference profile. Both bias corrections were estimated with an uncertainty of 0.03 %.

Because of the higher signal-to-noise ratios observed for cavity decays in data set B, owing to the improved laser linewidth and higher output power of the ECDL, the background noise was reduced by comparison to data set A and an etalon of period 2.82 GHz became apparent. The phase and amplitude of this etalon were then also floated.

The integrated spectral areas $A(T)$ were normalized to a reference temperature of $T_{\text{ref}} = 296$ K using the appropriate calculated total partition function $Q(T)$ [19, 27]:

$$A_{296} = A(T) \frac{T}{T_{\text{ref}}} \frac{Q(T)}{Q(T_{\text{ref}})} \exp\left[-\frac{hcE''}{k}\left(\frac{1}{T_{\text{ref}}} - \frac{1}{T}\right)\right], \tag{3}$$

where $E''$ are the lower-state vibrational energies, $h$ is the Planck constant, and $k$ is the Boltzmann constant.

Line intensities were obtained as the slope of a weighted linear regression of $A_{296}$ vs. number density $n_{\text{CO}_2} = \chi_{\text{CO}_2} p/(kT)$, where $p$ is the sample pressure and $T$ is the sample temperature. The intercept of this regression was fixed to zero. Figure 4 shows a representative example from data set B. All replicates were included simultaneously in the linear regression, and the weighting factors $w_i$ were derived from the fit quality factor defined earlier as $w_i = 1/\sigma_i^2$, with $\sigma_i = A_{296}/\text{QF}$.

Spectral interferences from weak $CO_2$ transitions were simulated and fixed during the fitting of the target $CO_2$ transitions using Voigt line shapes and known reference data [19]. Two transitions (R16 and R28) were also affected by spectral interference from water ($H_2O$). As a representative example from data set B, the intensity of line R16 (4866.264948 cm$^{-1}$) was corrected for the contribution of an underlying water line at 4866.25151 cm$^{-1}$, which was fully hidden in the spectra. The water mole fraction in the cell was estimated to be $\chi_{\text{H}_2\text{O}} = 26 \pm 5$ ppm, based on observation of another isolated water transition at 4874.27767 cm$^{-1}$, visible in the spectra about 5 GHz from the center of line R28. The intensity of line R28 was also corrected in this way. These corrections amounted to a relative change of −0.35(7) % for line R16 and +0.18(3) % for line R28.

While no further line-by-line corrections were required, we did perform a line-by-line analysis of the relative uncertainty contribution from weak $CO_2$ spectral interferences. For each affected line, this uncertainty in the retrieved area for the target $CO_2$ line was estimated as the relative change in its fitted, when the fixed intensity of the weak interfering line was modified by an amount equal to its tabulated uncertainty in HITRAN2016 (2 % to 5 %) [19]. These line-by-line systematic uncertainty estimates from known spectral interferences are tabulated later in the paper (see Section 5.1.3, Table 4).

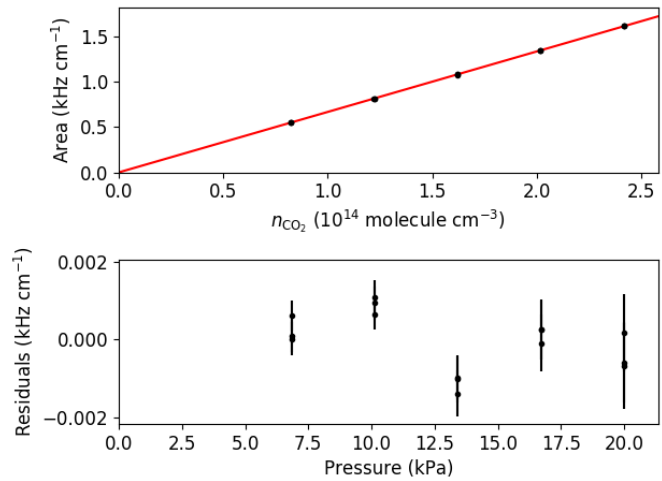

**Fig. 4.** Linear regression of spectral area vs. $CO_2$ number density, for line R24 recorded in data set B. The slope of a weighted linear regression yielded the line intensity S = 2.2220 × $10^{-22}$ cm molecule$^{-1}$ ± 5 × $10^{-26}$ cm molecule$^{-1}$. Residuals of the linear regression are also shown, with error bars corresponding to the standard uncertainty estimated as being inversely proportional to the fit quality factor.

## 4. Initial uncertainty budget

Table 2 summarizes the type B (systematic) uncertainty components common to each measurement. The remainder of this section discusses estimates of the individual type A uncertainties ($\sigma_{m,A}$, long-term reproducibility) for the purposes of performing a properly weighted least-squares analysis of the vibrational band intensity in Section 5.1.

**Table 2**
Type B (systematic) uncertainty budget for intensities ($S$) as measured by frequency-stabilized cavity ring-down spectroscopy.

| Parameter | $u_{r,S}^B$ (%) | Description |
|---:|:---:|:---|
| $\chi_{CO_2}$ | 0.04 | Sample mole fraction |
| $b_0, b_1, b_2$ | 0.04 | Digitizer nonideality |
| $g(\tilde{\nu})$ | 0.03 | Choice of line profile |
| $\nu_{FSR}$ | 0.02 | Cavity free spectral range |
| $w_m$ | 0.02 | Choice of weighting factors[a] |
| $p$ | 0.02 | Pressure and non-ideal gas effects |
| $T$ | 0.01 | Transition-dependent $T$ correction |
| $\chi_{626}$ | 0.007 | Isotopic abundance of $^{12}C^{16}O_2$ |
| **$u_{r,S}^B$** | **0.07** | **Combined type B uncertainty** |

[a] Weighting factors for vibrational band intensity fits, see Section 5.1.

Because of interferences from nearby water transitions as noted in Section 3, the R16 and R28 lines were not included in either the type A uncertainty evaluation presented here, or in the subsequent band fitting analysis discussed in Section 5.1. Here, we begin with a plot in Fig. 5 of the short-term type A fit uncertainties ($\sigma_{m,A(ST)}$) that resulted from individual fits of $A_{296}$ vs. $n_{CO_2}$ (e.g., Fig. 4). Data set A is plotted as red squares vs. $S$, and data set B is plotted as blue open circles.

Absent a robust set of repeated measurements at each line, we chose to estimate the long-term reproducibility [28] as follows, leveraging the fact that data sets A and B are a two-point test of long-term reproducibility. In Section 5.1, we will discuss two models for predicting individual line intensities ($S$) from an integrated band intensity ($S_{band}$). By fitting the models in Section 5.1 to both data sets, we assume that a reduced chi-squared ($\chi_\nu^2$) goodness-of-fit parameter greater than unity can be used to rescale the presumed under-estimation of the combined line-by-line variances $\sigma_{m,A(ST)}^2 + \sigma_{m,B}^2$, where $\sigma_{m,B}^2$ are the type B variances due to spectral interferences. Therefore, we write

$$\sigma_m^2 = \chi_\nu^2 \left\{ \left[ f_{\sigma_{m,A(ST)}} \right]^2 + \sigma_{m,B}^2 \right\} \quad (4)$$

for $\chi_\nu^2 > 1$, where $f_{\sigma_{m,A(ST)}}$ is a smooth function of $S$ which yields predicted values of $\sigma_{m,A(ST)}$. The purpose of the smoothing functions is to minimize the influence of large deviations in $\sigma_{m,A(ST)}$ that may arise from a low measurement sample size. The smoothing functions $f_{\sigma_{m,A(ST)}}$ are plotted as solid lines in Fig. 5 and assume a linear relationship between $\sigma_{m,A(ST)}$ and $S$.

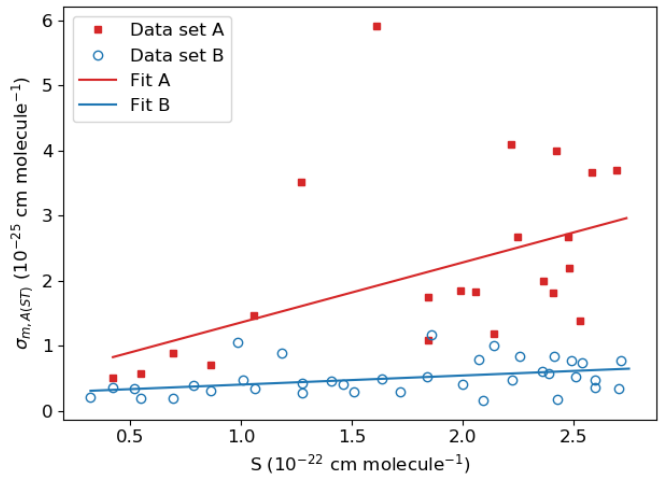

**Fig. 5.** Measured (symbols) and smoothed (solid lines) values for short-term type A measurement precision, $\sigma_{m,A(ST)}$ plotted vs. line intensities, $S$. Data set A is shown as full red squares, and data set B is shown as blue open circles.

The smoothing functions $f_{\sigma_{m,A(ST)}}$ aim to extract an estimate of the type-A reproducibility only. As seen in Fig. 5, several values of $\sigma_{m,A(ST)}$ appear to be outliers with relatively large short-term type-A uncertainties, especially for data set A (red squares). These outliers are attributed to the influence of spectral interferences (namely weak $CO_2$ and $H_2O$ lines), and we excluded data points for transitions which have direct overlap with known $H_2O$ lines (R16 and R28). To avoid biasing $f_{\sigma_{m,A(ST)}}$, we weighted our fits of the smoothing functions by the normalized values $1 - u_{r,S}^{m,B}/(\max u_{r,S}^{m,B})$. Consequently, transitions with the largest line-by-line type-B uncertainties attributed to spectral interferences were given a weight of zero and transitions uninfluenced by spectral interferences were given a weight of one. All other weights were linearly interpolated between the values of zero and one based on their relative systematic uncertainties $u_{r,S}^{m,B} = \sigma_{m,B}/S$.

## 5. Results and discussion

### 5.1. The 20013 ← 00001 integrated band intensity for $^{12}C^{16}O_2$

In this work, we measured the intensities of 39 unique $CO_2$ transitions with lower state rotational quantum numbers of $J'' \leq 44$ using several tunable lasers near $\lambda$ = 2.06 µm. With relatively broadband data, we were therefore able to determine an accurate integrated band intensity ($S_{\text{band}}$) for the 20013 ← 00001 band of $^{12}C^{16}O_2$ ($S_{\text{band}}^{626}$). While the fitting of vibrational band parameters is commonly done when very broadband spectral data is available (e.g., with Fourier transform spectroscopy [29]), it is less common for laser diode spectroscopy because of the normally limited number of accessible transitions. Here we overcame the limitations of tunability in continuous-wave laser spectroscopy by brute force: first by aligning four DFB lasers and then again by the introduction of a narrow linewidth ECDL. The result was a tabulation of 63 intensity measurements of 39 transitions with relative combined standard uncertainties ($u_{r,S}$) ranging from 0.13 % to 1.6 % (median value of $\tilde{u}_{r,S}$ = 0.26 %).

In this section, we apply and discuss two models for the individual intensities in order to determine $S_{\text{band}}$. In doing so, we show that the standard rotation-vibration coupling correction factor (i.e., the Herman-Wallis factor [30-33]) normally applied to the one-dimensional quantum mechanical model for $S$ are not required when we instead compare directly with the $J''$ dependence predicted by a fully *ab initio* dipole moment surface (DMS) [13]. For transitions with $J'' \leq 44$, we show that existing *ab initio* DMS calculations are as accurate as experiment in capturing the $J''$ dependence of $S$ for this combination-overtone band of a linear polyatomic molecule. Furthermore, subtle differences in the $J''$ dependent models elude to secondary perturbations beyond rotation-vibration coupling which are captured by the *ab initio* DMS. We also show that $S_{\text{band}}$ can be fitted equally well by either a properly normalized spectroscopic model or by applying a simple scaling of intensities from the *ab initio* DMS.

As introduced in Section 4, the proper choice of weighting factors for the band-wide fits is the inverse of the true type A variances associated with long-term reproducibility at each measurement point (i.e., each transition), $w_m = 1/\sigma_{m,A}^2$ [28]. However, line-by-line systematic contributions from spectral interferences have also been identified. Therefore, the chosen weighting factors are inversely proportional to the sum of type A (reproducibility) and type B (spectral interference, see Section 3) variances (e.g., $\sigma_m^2 = \sigma_{m,A}^2 + \sigma_{m,B}^2$). Values of $\sigma_{m,A}^2$ were estimated using Eq. (4) in Section 4, the smoothed functions $f_{\sigma_{m,A(ST)}}$ plotted in Fig. 5, and the values of $\chi_\nu^2$ calculated from a fitted model. To simplify the weighted least-squares regressions, we chose to use the normalized weighting factors

$$w_m^N = \frac{w_m}{\sum_m w_m} = \frac{1}{\sigma_m^2} \frac{1}{\sum_m \frac{1}{\sigma_m^2}}. \tag{5}$$

Critically, the value of $\chi_\nu^2$ is not required to calculate the normalized weighting factors in Eq. (5), and only the smoothed functions $f_{\sigma_{m,A(ST)}}$ plotted in Fig. 5 calculated from the observed short-term trends in fit precision and the systematic contributions from spectral interferences $\sigma_{m,B}^2$ shown in Eq. (4) must be known *a priori*.

### 5.1.1. Standard spectroscopic model

The individual rotational-vibrational line intensities of a linear molecule can be calculated from the following spectroscopic model [34] with its origins in quantum mechanics [35]:

$$S_m = \frac{S_V \chi_{iso}}{g \tilde{\nu}_0 Q_R(T)} \tilde{\nu}_m L_m F_m \left[1 - \exp\left(-\frac{hc\tilde{\nu}_m}{kT}\right)\right] \exp\left(-\frac{hcE_m''}{kT}\right), \tag{6}$$

where $S_V$ is the vibrational band intensity without perturbation from rotation-vibration coupling, $\chi_{iso}$ is the reference isotopic abundance (e.g., $\chi_{626,HT}$), $g$ is the degeneracy of the rotational-vibrational state, $\tilde{\nu}_0$ is the origin frequency of the vibrational band in cm$^{-1}$, $Q_R(T)$ is the rotational partition function, $\tilde{\nu}_m$ are the individual transition frequencies in cm$^{-1}$, $L_m$ are the Hönl-London factors, $F_m$ are the Herman-Wallis factors, and $E_m''$ are the lower state rotational energies in dimensions of wave number. For a more detailed discussion of Eq. (6) and its constituent terms, see Ref. [29].

For the 20013 ← 00001 $^{12}C^{16}O_2$ band, we calculate from Section 4.1 of [29] that $g = 1$ and $L_m = |m|$, where $m = -J''$ for the P-branch ($\Delta J = J' - J'' = -1$) and $m = J'' + 1$ for the R-branch ($\Delta J = +1$), and $J'$ is the upper state rotational quantum number. For the form of the Herman-Wallis factor, we choose

$$F_m = (1 + a_1 m + a_2 m^2 + a_3 m^3)^2, \tag{7}$$

where $a_1$, $a_2$, and $a_3$ are fitted coefficients [along with $S_V$ in Eq. (6)]. In applying Eq. (6), the following terms were held constant: $\tilde{\nu}_0 = 4854.447$ cm$^{-1}$ [13], $T = T_{\text{ref}} = 296$ K, and $Q_R(T_{\text{ref}}) = 263.87063$ [19, 27]. Values for $\chi_{\text{iso}} = \chi_{626,\text{HT}}$, $E''$ and $\tilde{\nu}$ were taken from HITRAN2016 [19].

We call attention here to the subtle difference between a vibrational band intensity ($S_V$) and the integrated band intensity, $S_{\text{band}} = \sum_m S_m$ [10]. If the Herman-Wallis factors in Eq. (6) are neglected (i.e., $F_m = 1$), then $S_{\text{band}} = S_V$. However, once rotation-vibration coupling is introduced the values of $S_V$ and $S_{\text{band}}$ diverge by consequence of their respective definitions. This is reconciled here by the introduction of $F_0$, or the Herman-Wallis normalization factor, where $S_{\text{band}} = F_0 S_V$.

**Table 3**
Summary of the fitted 20013 ← 00001 $^{12}C^{16}O_2$ vibrational band intensity parameters and reduced chi-squared values. Standard uncertainties are shown in parenthesis and represent the type A fit precisions retrieved from each model.

| Parameter | Units | Spectroscopic model | *ab initio* model |
|---|---|---|---|
| $S_{\text{band}}^{626}$ | cm molecule$^{-1}$ | 7.182(2)×10$^{-21}$ | 7.1830(17)×10$^{-21}$ |
| $a_1$ | | 1.96(11)×10$^{-4}$ | |
| $a_2$ | | 2.98(3)×10$^{-5}$ | |
| $a_3$ | | 16(14)×10$^{-8}$ | |
| $F_0$ | | 1.0323 | |
| $\chi_\nu^2$ | | 14.6 | 19.8 |

The fitted values of $a_1$, $a_2$ and $a_3$ are listed in Table 3, along with the normalization factor $F_0$. Also listed is the isotopic-abundance-normalized integrated band intensity, $S_{\text{band}}^{626} = F_0 S_V \chi_{626,\text{HT}}$. Following a first fitting iteration with $F_0 = 1$, the normalization factor $F_0$ was calculated imposing the constraint $S_{\text{band}}^{626} = \sum_m S_m$ and subsequently fixed for a second and final fitting iteration. The full numerical normalization factor $F_0$ is written in Eq. (8)

$$F_0 = \frac{1}{\tilde{v}_0 Q_R(T)} \sum_m \tilde{v}_m L_m F_m \left[1 - \exp\left(\frac{hc\tilde{v}_m}{kT}\right)\right] \exp\left(\frac{-hcE_m''}{kT}\right), \qquad (8)$$

where the summation over $m$ included 109 transitions with $J'' < 110$, equal to the total number of 20013 ← 00001 $^{12}$C$^{16}$O$_2$ transitions in HITRAN2016 [19].

### 5.1.2. Application of an *ab initio* dipole moment surface

Intensities reported here for the 20013 ← 00001 $^{12}$C$^{16}$O$_2$ band near $\lambda$ = 2.06 μm were also compared to those available in the HITRAN2016 database [19], originally calculated by Zak et al. using an accurate *ab initio* DMS [13]. While the spectroscopic model discussed in the preceding Section 5.1.1 has its origins in quantum chemistry modeling, several basic assumptions are required to calculate the individual intensities using Eq. (6). As early evidence of its limitations, the $F_m$-style correction factors were introduced originally by Herman and Wallis [30] to account for perturbations to the rotational intensities due to rotation-vibration coupling. However, with a fully *ab initio* DMS available for CO$_2$ and its isotopologues [13, 36, 37], we can simply forego the spectroscopic model and compare directly with intensities from the multidimensional *ab initio* DMS.

Plotted in Fig. 6 are the line-by-line values reported herein ($S_{\text{NIST}}$) vs. the corresponding values in HITRAN2016 ($S_{\text{HT}}$) [19] calculated from the *ab initio* DMS [13]. A weighted least-squares fit of the parameter $\beta$, where $S_{\text{NIST}} = \beta S_{\text{HT}}$, yielded the black line in Fig. 6 and the scaling parameter $\beta$ = 1.0069 with a relative uncertainty of $u_{r,\beta}$ = 0.02 %. The integrated band intensity $S_{\text{band}}^{626}$ was then calculated using Eq. (9)

$$S_{\text{band}}^{626} = \beta \sum S_{\text{HT}}, \qquad (9)$$

where the summation was again over 109 transitions with $J'' < 110$. (We confirmed numerically that $\sum S_{\text{HT}}$ was equal to the integrated band intensity reported in Table 1 of [13].) Therefore, in addition to the spectroscopic model discussed in Section 5.1.1, we also report in Table 3 of this work a fitted value for $S_{\text{band}}^{626}$ which assumed a $J''$ dependence from the *ab initio* DMS: $S_{\text{band}}^{626} = 7.1830 \times 10^{-21}$ cm molecule$^{-1}$ ($u_{r,S_{\text{band}}^{626}}^{A} = 0.02$ %).

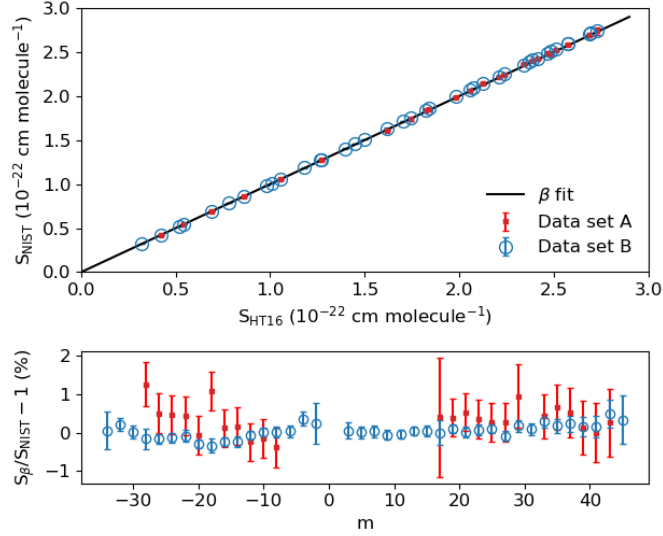

**Fig. 6.** Weighted regression of $S_{\text{NIST}}$ vs. $S_{\text{HT16}}$ is shown in the top panel, where $S_{\text{NIST}}$ comprises both data sets A (red squares) and B (blue circles). The fitted scaling factor (solid black line) was $\beta = 1.0069 \pm 0.0002$ (relative uncertainty of 0.02 %) — meaning that the NIST values for the 20013 ← 00001 $^{12}C^{16}O_2$ band are 0.69 % larger than those tabulated in HITRAN2016 [19]. Relative fit residuals for both data sets A and B are plotted in the bottom panel. Error bars represent $1\sigma$ type A uncertainty in $A_{296}$ vs. $n_{CO_2}$ (see Table 4).

### 5.1.3. Reported line-by-line intensities

Here we report the line-by-line intensities measured in data sets A and B, as well as the intensities that resulted from weighted fits of the combined data sets to a spectroscopic model (Section 5.1.1) and an *ab initio* DMS (Section 5.1.2). The results are summarized in Table 4.

**Table 4**
Summary of line-by-line transition intensities for the 20013 ← 00001 $^{12}C^{16}O_2$ vibrational band. Line-by-line relative standard uncertainties (type A reproducibility and type B spectral interferences only) from the weighted regressions of $A_{296}$ vs. $n_{CO_2}$ ranged from 0.6 % to 1.6 % (data set A) and from 0.10 % to 0.6 % (data set B).

-21-

| Transition | m | Spectral interference | Data set A | | Data set B | | Spectroscopic model | | *ab initio* model | |
|---|---|---|---|---|---|---|---|---|---|---|
| | | $u_{r,S}^{m,B}$ | $S \times 10^{22}$ | $u_{r,S}^m$ | $S \times 10^{22}$ | $u_{r,S}^m$ | $S \times 10^{22}$ | $u_{r,S}^A$ | $S \times 10^{22}$ | $u_{r,S}^A$ |
| | | % | cm mlc$^{-1}$ | % | cm mlc$^{-1}$ | % | cm mlc$^{-1}$ | % | cm mlc$^{-1}$ | % |
| P34 | −34 | 0.1 | | | 0.984 | 0.5 | 0.9848 | 0.16 | 0.9846 | 0.02 |
| P32 | −32 | 0.01 | | | 1.185 | 0.17 | 1.1873 | 0.14 | 1.1872 | 0.02 |
| P30 | −30 | 0.02 | | | 1.404 | 0.17 | 1.4050 | 0.13 | 1.4047 | 0.02 |
| P28 | −28 | 0.05 | 1.610 | 0.6 | 1.633 | 0.3 | 1.6311 | 0.11 | 1.6302 | 0.02 |
| P26 | −26 | 0 | 1.846 | 0.5 | 1.858 | 0.13 | 1.8564 | 0.10 | 1.8548 | 0.02 |
| P24 | −24 | 0.01 | 2.058 | 0.5 | 2.071 | 0.13 | 2.0695 | 0.09 | 2.0682 | 0.02 |
| P22 | −22 | 0.02 | 2.246 | 0.5 | 2.257 | 0.14 | 2.2577 | 0.08 | 2.2555 | 0.02 |
| P20 | −20 | 0 | 2.406 | 0.5 | 2.412 | 0.11 | 2.4069 | 0.07 | 2.4046 | 0.02 |
| P18 | −18 | 0.03 | 2.475 | 0.5 | 2.510 | 0.17 | 2.5033 | 0.06 | 2.5012 | 0.02 |
| P16 | −16 | 0 | 2.528 | 0.5 | 2.537 | 0.11 | 2.5338 | 0.06 | 2.5314 | 0.02 |
| P14 | −14 | 0.02 | 2.481 | 0.5 | 2.491 | 0.14 | 2.4877 | 0.05 | 2.4851 | 0.02 |
| P12 | −12 | 0.01 | 2.361 | 0.5 | 2.357 | 0.12 | 2.3577 | 0.05 | 2.3552 | 0.02 |
| P10 | −10 | 0.01 | 2.142 | 0.5 | 2.138 | 0.13 | 2.1407 | 0.04 | 2.1387 | 0.02 |
| P8 | −8 | 0.01 | 1.844 | 0.5 | 1.836 | 0.13 | 1.8387 | 0.04 | 1.8366 | 0.02 |
| P6 | −6 | 0.01 | | | 1.458 | 0.15 | 1.4590 | 0.04 | 1.4580 | 0.02 |
| P4 | −4 | 0.01 | | | 1.009 | 0.18 | 1.0140 | 0.03 | 1.0130 | 0.02 |
| P2 | −2 | 0.1 | | | 0.5193 | 0.5 | 0.5209 | 0.03 | 0.52048 | 0.02 |
| R0 | 1 | | | | | | | | 0.26372 | 0.02 |
| R2 | 3 | 0 | | | 0.7830 | 0.2 | 0.7837 | 0.04 | 0.7834 | 0.02 |
| R4 | 5 | 0 | | | 1.274 | 0.15 | 1.2745 | 0.04 | 1.2738 | 0.02 |
| R6 | 7 | 0.02 | | | 1.715 | 0.16 | 1.7157 | 0.04 | 1.7158 | 0.02 |
| R8 | 9 | 0.01 | | | 2.092 | 0.13 | 2.0902 | 0.05 | 2.0904 | 0.02 |
| R10 | 11 | 0 | | | 2.386 | 0.11 | 2.3853 | 0.05 | 2.3854 | 0.02 |
| R12 | 13 | 0.01 | | | 2.594 | 0.12 | 2.5937 | 0.06 | 2.5949 | 0.02 |
| R14 | 15 | 0.02 | | | 2.713 | 0.14 | 2.7134 | 0.07 | 2.7147 | 0.02 |
| R16[a] | 17 | 0.07 | 2.74 | 1.6 | 2.749 | 0.3 | 2.7474 | 0.08 | 2.7489 | 0.02 |
| R18 | 19 | 0.01 | 2.695 | 0.5 | 2.703 | 0.11 | 2.703 | 0.09 | 2.7056 | 0.02 |
| R20 | 21 | 0.02 | 2.581 | 0.5 | 2.594 | 0.14 | 2.592 | 0.10 | 2.5949 | 0.02 |
| R22 | 23 | 0.02 | 2.421 | 0.5 | 2.428 | 0.14 | 2.427 | 0.12 | 2.4297 | 0.02 |
| R24 | 25 | 0 | 2.219 | 0.5 | 2.223 | 0.12 | 2.222 | 0.13 | 2.2253 | 0.02 |
| R26 | 27 | 0.01 | 1.989 | 0.5 | 1.997 | 0.13 | 1.993 | 0.15 | 1.9947 | 0.02 |
| R28[a] | 29 | 0.03 | 1.737 | 0.8 | 1.750 | 0.15 | 1.751 | 0.17 | 1.7531 | 0.02 |
| R30 | 31 | 0.02 | | | 1.509 | 0.17 | 1.509 | 0.19 | 1.5104 | 0.02 |
| R32 | 33 | 0.02 | 1.272 | 0.6 | 1.274 | 0.18 | 1.276 | 0.2 | 1.2778 | 0.02 |
| R34 | 35 | 0.01 | 1.054 | 0.6 | 1.059 | 0.18 | 1.059 | 0.2 | 1.0613 | 0.02 |
| R36 | 37 | 0.01 | 0.861 | 0.6 | 0.8631 | 0.2 | 0.863 | 0.3 | 0.8651 | 0.02 |
| R38 | 39 | 0.02 | 0.692 | 0.7 | 0.6920 | 0.2 | 0.6917 | 0.3 | 0.6931 | 0.02 |

| | | | | | | | | | |
|---|---|---|---|---|---|---|---|---|---|
| R40 | 41 | 0 | 0.546 | 0.8 | 0.5448 | 0.3 | 0.5447 | 0.3 | 0.54576 | 0.02 |
| R42 | 43 | 0.02 | 0.421 | 0.9 | 0.4204 | 0.4 | 0.4216 | 0.4 | 0.42251 | 0.02 |
| R44 | 45 | 0.1 | | | 0.3205 | 0.6 | 0.3209 | 0.4 | 0.32161 | 0.02 |

[a] Overlap with known spectral interference from water. Transition not included in band intensity analysis.

In Table 4, the $u^A_{r,S}$ values associated with the fitted $\beta$ parameter ("*ab initio* model") are all identical, with no $m$ dependence in the scaling procedure. In contrast, the $u^A_{r,S}$ for the "spectroscopic model" have a noticeable $m$ dependence caused by propagation of uncertainty in the fitted $F_m$ coefficients comprising Eq. (7). Relative values of $S$ calculated from both band fitting models are plotted vs. $m$ in Fig. 7. In Fig. 7, the shaded area represents the uncertainty in this relative difference as a quadrature sum of $u^A_{r,S}$ in Table 4 (both models). Small, systematic differences are observed when comparing the spectroscopic model and *ab initio* DMS model. However, the differences are on the order of the combined uncertainties plotted in Fig. 7.

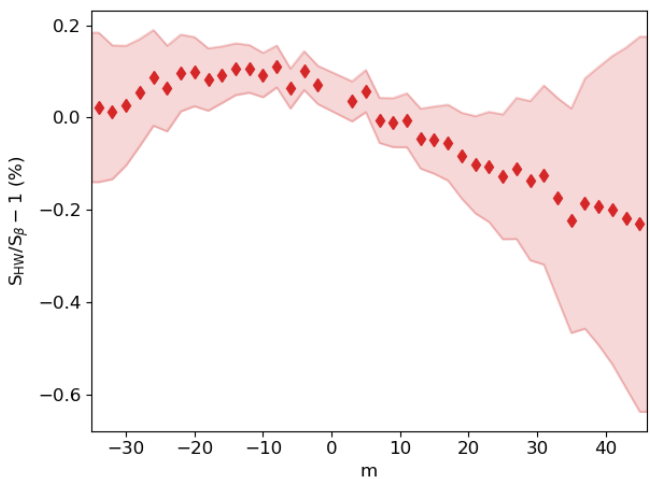

**Fig. 7.** Plotted vs. $m$ is the relative difference between the spectroscopic model for intensities which included Herman-Wallis factors ($S_{\mathrm{HW}}$) and a scaled version of the ab initio DMS intensities ($S_\beta$). The shaded region represents the quadrature sum of type-A fit precision from each model.

Given that the $J''$ dependence of the *ab initio* DMS [as well as that of the underlying potential energy surface (PES)] should capture phenomena beyond just rotation-vibration coupling, we choose to report the $\beta$ parameter scaling of the *ab initio* DMS as our experimentally derived values for $S_{NIST}$: $S_{band}^{626} = 7.183 \times 10^{-21}$ cm molecule$^{-1}$ $\pm$ 6 $\times$ 10$^{-24}$ cm molecule$^{-1}$. The relative combined standard uncertainty ($u_{r,S}$ = 0.08 %) resulted from the quadrature sum of the type A fit precision ($u_{r,S}^A$ = 0.02 %) from Section 5.1.2 shown in Table 4 of this section and the relative combined type B (systematic) standard uncertainty ($u_{r,S}^B$ = 0.07 %) shown in Table 2 of Section 4. A final line list inclusive of our total measurement range is available as supplemental material. From $S_{band}^{626}$, we calculate the vibrational band intensity of the 20013 ← 00001 CO$_2$ band to be $S_V = S_{band}^{626}/(F_0 \chi_{626,HT}) = 7.070 \times 10^{-21}$ cm molecule$^{-1}$ $\pm$ 6 $\times$ 10$^{-24}$ cm molecule$^{-1}$.

## 5.2. Final uncertainty evaluation

As detailed in Section 5.1.3, we report a line list for the 20013 ← 00001 band of $^{12}C^{16}O_2$ as a scaled version of the intensities calculated from the *ab initio* DMS of Zak et al. [13]. The scaling factor, $\beta$ = 1.0069, leads to an integrated band intensity of 7.183 $\times$ 10$^{-21}$ cm molecule$^{-1}$ $\pm$ 6 $\times$ 10$^{-24}$ cm molecule$^{-1}$. Because the *ab initio* DMS intensities display a $J''$ dependence which matches our experimental observations, we fit only the scaling factor $\beta$. Therefore, the relative combined uncertainties for all transitions reported in our line list are equal to the quadrature sum of the type A (fit precision on $\beta$) and type B (see Table 2) relative uncertainties, $u_{r,S}$ = 0.08 %. The consistency in $u_{r,S}$ vs. $J''$ (or $m$, where $m = -J''$ for the P-branch and $m = J'' + 1$ for the R-branch) is a consequence of only fitting one parameter. All of our measured intensities weighted by the inverse of their combined variances contribute to the value of $\beta$. Therefore, scatter in these data about the *ab initio* model provides a measure of the long-term measurement reproducibility, whereas uncertainty in the weighted mean value of $\beta$ affects the accuracy of the measured band intensity.

The line-by-line type A uncertainties reported in Table 4 for the spectroscopic model fit are very different and display a clear $m$ dependence. This is not surprising and is a consequence of floating three Herman-Wallis coefficients ($a_1$, $a_2$, and $a_3$) with their own levels of fit precision (and covariance). The values of $u_{r,S}^A = \sigma_S^A/S$ reported for the spectroscopic model in Table 4, where $\sigma_S^A$ is the type A standard uncertainty, were calculated using the full covariance matrix and Eq. (10),

$$(\sigma_S^A)^2 = \sum_{i=1}^{N} \left(\frac{\partial f}{\partial x_i}\right) \sigma^2(x_i) + 2 \sum_{i=1}^{N-1} \sum_{j=i+1}^{N} \frac{\partial f}{\partial x_i} \frac{\partial f}{\partial x_j} \sigma(x_i, x_j). \tag{10}$$

Above, $f$ represents the model in Eq. (6), $x_i$ are the floated parameters (quantity $N$ = 4), $\sigma^2(x_i)$ are the diagonal elements of the covariance matrix, and $\sigma(x_i, x_j)$ are the off-diagonal elements (co-variances) [38]. Analytical expressions for the partial derivatives enabled accurate calculation from the estimated covariance matrix.

### 5.3. *Constraining the Herman-Wallis expansion for $CO_2$*

For the nearby 20012 ← 00001 band of $^{12}C^{16}O_2$ at $\lambda \approx 2.0$ μm, Casa et al. reported a constraint on the accuracy of the Herman-Wallis expansion of 0.5 % [39]. However, the constraint was limited by the number of intensity measurements reported within only the R-branch, up to $m$ = 19. Further, subsequent measurements [14, 40, 41] and *ab initio* calculations [13] revealed systematic shifts in the original Casa et al. data attributed to the choice of line shape. Given the large number of intensities reported here, from $m = -34$ to $m = 45$, and our comprehensive, low uncertainty budget, we revisit the concept of constraining the accuracy of the Herman-Wallis expansion for $CO_2$.

First, we note that the fitted values of $S_{band}^{626}$ retrieved from both the spectroscopic model and the *ab initio* model agreed to within their reported precisions of about 0.02 % to 0.03 % (see Table 3). While $S_{band}^{626}$ scales the line-by-line intensities [see Eq. (6)], we can further analyze the $m$-dependence of the spectroscopic model by focusing on just the Herman-Wallis expansion. Shown in Fig. 8 are the weighted residuals from the fitted spectroscopic model with 3$^{rd}$-order Herman-Wallis expansion [see Eq. (7)], relative to the measured intensities. Using the weighted residuals, we calculate the root mean squared deviation $\hat{\sigma}$ from the weighted fit as:

$$\hat{\sigma} = \sqrt{\sum_m w_m^N \left[\frac{(S_m - S_m^{HW})}{S_m}\right]^2} \tag{11}$$

where $w_m^N$ are again the normalized weights ($\sum_m w_m^N = 1$), $S_m$ are the reported intensities measured here, and $S_m^{HW}$ are the intensities fitted by the spectroscopic model. The result is a root mean squared deviation of $\hat{\sigma} = 0.18$ %, only a factor of 2.25 larger than our combined relative uncertainty of 0.08 %. Therefore, we postulate based on our experimental observations that the Herman-Wallis expansion can accurately reproduce the $m$-dependence of the 20013 ← 00001 band intensities to within $\hat{\sigma} = 0.18$ %.

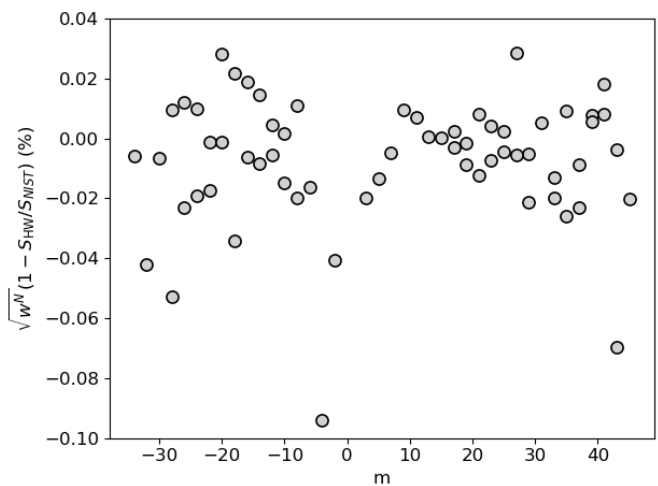

**Fig. 8.** Plotted vs. $m$ are the weighted residuals from the fitted spectroscopic model, relative to the measured values of $S_{NIST}$. Visual inspection of the weighted residuals suggests that the spectroscopic model with Herman-Wallis expansion is sufficient to capture $m$-dependent trends in the experimental data. The root mean squared deviation for the spectroscopic model with Herman-Wallis expansion is $\hat{\sigma} = 0.18$ %.

Our rigorous evaluation of the accuracy of the Herman-Wallis expansion is further strengthened by the availability of highly accurate *ab initio* calculations. The root mean squared deviation from the weighted fit to the *ab initio* model was similar, with a value of $\hat{\sigma} = 0.23$ %. Collectively, our observations and analysis comprise the most rigorous comparison of spectroscopic models, *ab initio* calculations, and low-uncertainty experiments for $CO_2$ to date — fulfilling the relative uncertainty criteria of ≤0.1 % for reference line intensities required to achieve precision remote sensing of the Earth's atmosphere. Furthermore, as enabled by our highly accurate FS-CRDS measurements, we explicitly show that $CO_2$ intensities predicted by a fitted spectroscopic model with Herman-Wallis expansion are consistent with a simply rescaling of the integrated band intensity calculated by an *ab initio* dipole moment surface.

### 5.4. Comparisons with the literature: additional CO₂ line lists

Figure 9 presents a comparison of our results with several databases and line lists available in the literature: intensities from HITRAN2012 [29, 42], Benner et al. [10], Ames-2016 [43], and the carbon dioxide spectroscopic databank, CDSD-296 [44] are plotted relative to our scaled *ab initio* model. The Fourier transform spectroscopy (FTS) line list of Benner et al. [10] has a $J''$ dependence which is reproduced well by our data and by the *ab initio* DMS intensities over the range of $J'' \leq 44$. The Ames-2016 line list of Huang et al. [43], also an *ab initio* DMS calculation, displays a similar $J''$ dependence. Significantly, the summary plot in Fig. 9 therefore shows good agreement between two independent experimental measurements (FS-CRDS and FTS) as well as two nearly independent theoretical approaches (University College London, UCL, and Ames-2016) on the $J''$ dependence of the intensities within the $20013 \leftarrow 00001$ $^{12}C^{16}O_2$ vibrational band.

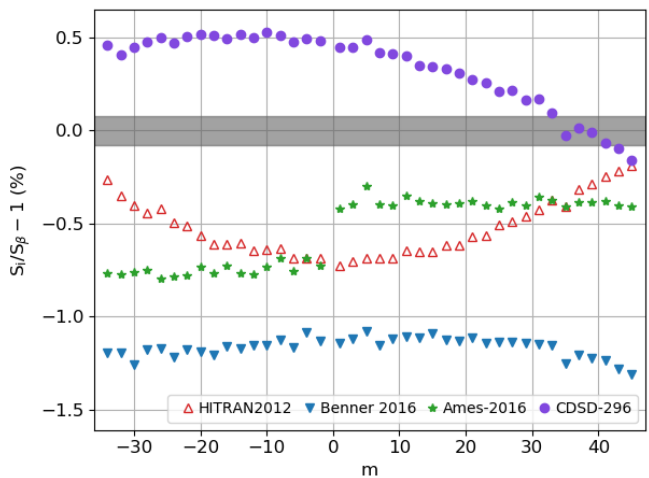

**Fig. 9.** Comparison with reference databases. Plotted are relative differences $S_i/S_\beta - 1$ where $S_\beta = \beta S_{HT}$ are the scaled *ab initio* DMS intensities [13]. The shaded region represents our relative combined standard uncertainty of 0.08 %. The relative standard uncertainties for each line list (not shown) are: HITRAN2012, 1 % [29, 42]; Benner 2016, 1 % [10]; Ames-2016, 1–3 % [43]; CDSD-296, 1.6 % [44].

The line lists, however, do disagree significantly in the magnitude of the integrated band intensity, $S_{band}^{626}$. From the line lists we calculated the following average relative deviations from our scaled *ab initio* DMS line list: HITRAN2012: –0.52 % ± 0.15 %; Benner et al.: –1.16 % ± 0.05 %; Ames-2016: –0.54 % ± 0.18 %; CDSD-296: +0.33 % ± 0.20 %.

*5.5.    Comparisons with the literature:  other measurements*

Figure 10 shows an additional comparison of our scaled *ab initio* model with two publications from the CNRS group in Reims, France: a combined Fourier transform spectroscopy and tunable diode laser absorption spectroscopy (TDLAS) study by Régalia et al. [45], and 6 lines measured with laser diode absorption spectroscopy reported by Joly et al. [46]. The average agreement is quite good despite the large scatter found in the prior experiments. For the 2.06 μm band, Régalia et al. [45] performed multispectral fits using Voigt profiles to model absorption from single-pass sample cells containing high-purity $CO_2$. In Joly et al. [46] several transitions were revisited again using TDLAS and Voigt profiles but with a multipass absorption cell. While, on average, agreement with the NIST values in Fig. 10 is good, Voigt profiles have been shown numerically to underestimate molecular line areas by 1 % to 4 % [47]. Therefore, the agreement may be partly coincidental.

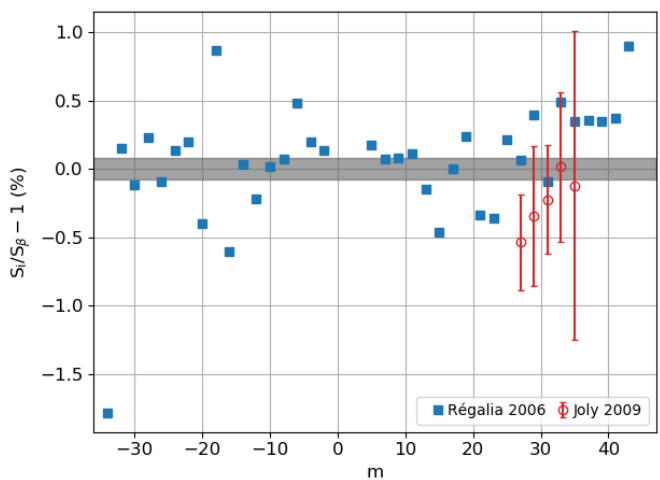

**Fig. 10.** Comparison with FTS [45] and laser diode absorption spectroscopy [46] measurements from CNRS. Plotted are relative differences $S_i/S_\beta - 1$ where $S_\beta = \beta S_{\text{HT}}$ are the scaled *ab initio* DMS intensities [13]. The shaded region represents our relative combined standard uncertainty of 0.08 %. The relative combined standard uncertainty for the intensities in Ref. [45] (not shown) is about 2 %.

## 6. Conclusions

Here we report accurate frequency-stabilized cavity ring-down spectroscopy measurements of the 20013 ← 00001 vibrational band of $^{12}C^{16}O_2$ near $\lambda$ = 2.06 μm. As one of the two CO$_2$ bands observed by the NASA Orbiting Carbon Observatory-2, the so-called "strong band" at $\lambda$ = 2.06 μm is important for space-based atmospheric measurements of carbon sources and sinks [48, 49] as well as various near-ground based regional observation platforms [50-52]. Our broadband measurements, performed using five different tunable continuous-wave lasers, were analyzed to yield fitted vibrational band intensities using two different models. Firstly, we fitted the measured values to the traditional spectroscopic model derived from fundamental assumptions of quantum mechanics, including Herman-Wallis factors with proper normalization. Secondly, we used our experimental values to scale the calculated *ab initio* intensities [13] found in HITRAN2016 [19]. This approach assumes that the *ab initio* dipole moment surface (DMS) of Zak et al. [13] accurately yields the $J''$ dependence of the line-by-line intensities.

To reproduce the experimentally observed transition intensities, we find that the traditional spectroscopic model with the inclusion of three normalized Herman-Wallis factors, performed equally as well as a single-parameter scaling of the *ab initio* DMS values. However, given that the *ab initio* DMS is more physically general than the Herman-Wallis spectroscopic model and requires only a single floated parameter to fit to our measurements, we chose the results of Ref. [13] as the best available representation of the true CO$_2$ DMS. To this end, we report an integrated band intensity for the 20013 ← 00001 vibrational band of $^{12}C^{16}O_2$ near $\lambda$ = 2.06 μm of $S_{band}^{626}$ = 7.183 × 10$^{-21}$ cm molecule$^{-1}$ ± 6 × 10$^{-24}$ cm molecule$^{-1}$ ($u_{r,S}$ = 0.08 %). This experimental band intensity, used in combination with the $J''$ dependence calculated from the *ab initio* DMS, allowed for a comparison of our accurate intensities with those available from other line lists and prior laser spectroscopy experiments.

Self-consistent and accurate line-by-line carbon dioxide reference data should encompass line intensities, line mixing coefficients and line shape parameters valid over a relevant range of temperature and pressure. These data are critical for further reduction of spectroscopic uncertainties in $XCO_2$ retrieval error budgets [3]. Thompson et al. [53] improved 2.06 μm band retrievals through the addition of an *ad hoc* continuum, which is used in current ABSCO v5.0. The need for an *ad hoc* continuum could come from the nearest-neighbor line mixing approximation [10] used in ABSCO v5.0, which does not account for coupling between all combinations of lines, specifically between the P and R branches [54]. Also, given that line mixing involves intensities of the interacting lines, our reported values strongly constrain line-mixing models that are fit to measured spectra for this band. Future work will include band-wide laser scans and frequency-stabilized cavity ring-down spectroscopy measurements that probe a pressure range where line mixing and non-Voigt profile line shape effects are important. We seek to include the current study with these new measurements to yield a self-consistent set of reference data.

## Acknowledgments

Funding:  This work was supported by the National Institute of Standards and Technology (NIST); the NIST Greenhouse Gas and Climate Science Program; and the National Aeronautics and Space Administration (NASA) [contract number NRA NNH17ZDA001N-OCO2].

D. Michelle Bailey (NIST) provided comments on the manuscript.

## Competing Interests

The authors declare no competing interests.